\documentclass[aps,pra,floatfix,showpacs,groupedaddress,notitlepage,twocolumn,
amsmath,amssymb]{revtex4}

\usepackage{graphicx}
\usepackage{bm}
\usepackage{subfigure}
\usepackage{braket}
\usepackage{times}

\newcommand{\ofr}{(\mathbf{r})}

\newcommand{\nR}{{n_\mathrm{R}}}
\newcommand{\nL}{{n_\mathrm{L}}}
\newcommand{\R}{\mathrm{R}}
\renewcommand{\L}{\mathrm{L}}

\renewcommand{\a}{{(\alpha)}}
\newcommand{\tot}{{\mathrm{tot}}}
\newcommand{\eff}{{\mathrm{eff}}}
\newcommand{\ER}{E_\mathrm{R}}
\newcommand{\as}{a_\mathrm{B}}
\newcommand{\asBF}{a_\mathrm{BF}}
\newcommand{\expect}[1]{\langle #1 \rangle}
\newcommand{\I}{\mathrm{I}}
\newcommand{\F}{\mathrm{F}}
\newcommand{\hb}[2]{\hat{b}_\text{#1}^{(#2)}}
\newcommand{\hbd}[2]{\hat{b}_\text{#1}^{(#2)\dagger}}
\newcommand{\hf}[2]{\hat{f}_\text{#1}^{(#2)}}
\newcommand{\hfd}[2]{\hat{f}_\text{#1}^{(#2)\dagger}}
\newcommand{\dr}{\int d^3{r}\ }

\newcommand{\B}{\mathrm{B}}
\renewcommand{\F}{\mathrm{F}}
\newcommand{\BF}{\mathrm{BF}}
\newcommand{\BB}{\mathrm{BB}}

\newcommand{\Rb}{^{87}\mathrm{Rb}}
\newcommand{\K}{^{40}\mathrm{K}} 
\newcommand{\abcd}{\alpha\beta\gamma\delta}
 
\newcommand{\fig}[1]{Fig.~\ref{#1}}

\begin{document}

\title{Density-induced processes in quantum gas mixtures in optical lattices} 

\author{Ole J\"urgensen}
\author{Klaus Sengstock}
\author{Dirk-S\"oren L\"uhmann}
\affiliation{%
 Institut f\"ur Laser-Physik, Universit\"at Hamburg, Luruper Chaussee 149, 22761 Hamburg, Germany
}%
 

\begin{abstract}
We show that off-site processes and multi-orbital physics have a crucial impact on the phase diagram of quantum gas mixtures in optical lattices. In particular, we discuss Bose-Fermi mixtures where the intra- and interspecies interactions induce competing density-induced hopping processes, the so-called bond-charge interactions. Furthermore, higher bands strongly influence tunneling and on-site interactions. We apply a multi-orbital interaction-induced dressing of the lowest band which leads to renormalized hopping processes. These corrections give rise to an extended Hubbard model with intrinsically occupation-dependent parameters. The resulting decrease of the tunneling competes with a decrease of the total on-site interaction energy both affecting the critical lattice depth of the superfluid to Mott insulator transition. In contrast to the standard Bose-Fermi-Hubbard model, we predict a large shift of the transition to shallower lattice depths with increasing Bose-Fermi attraction. The applied theoretical 
model allows an accurate prediction of the modified tunneling amplitudes and the critical lattice depth both recently observed experimentally.   
\end{abstract}

\pacs{03.75.Lm, 67.85.Pq, 71.10.Fd}

\maketitle

Quantum gas mixtures in optical lattices are well suited to study in detail interaction-induced effects in condensed matter. They allow for the investigation of systems with spin degree of freedom and with different species of particles that can even obey different quantum statistics.
In particular, the experimental realization of atomic mixtures of bosonic and fermionic particles (e.g., $^{87}$Rb-$^{40}$K) in optical lattices \cite{Gunter2006,Ospelkaus2006,Best2009,Heinze2011} triggered a vivid discussion on the role of inter- and intraspecies interactions. These experiments allow for the observation of the bosonic superfluid to Mott-insulator transition in the presence of fermionic atoms.
The prominent feature observed in all experiments is the decay of visibility and condensate fraction of the bosonic subsystem induced by the interaction with the fermionic atoms. Two possible explanations for this drop in bosonic coherence were proposed. First, the process of adiabatic heating while ramping the lattice has been suggested  \cite{Cramer2008,Cramer2011}. It is caused by different contributions of the atomic species to the total entropy and is therefore specific to the loading procedure of experiments with ultracold gases. Second, an interaction-induced dressing of tunneling and interaction processes has been found that causes a shift of the superfluid to Mott-insulator phase transition \cite{Luhmann2008b,Best2009,Lutchyn2009,Mering2011}. The latter effect corresponds to a necessary extension of the Hubbard model at zero temperature and is therefore fundamental for various lattice systems. The important role of interaction-induced processes in optical lattices is caused by the specific shape of 
the Wannier functions and the possibility of high filling factors.

The {\it standard} Bose-Fermi-Hubbard model \cite{Albus2003} is restricted to the lowest single-particle band and on-site interactions. Interestingly, it fails to describe interaction effects in boson-fermion mixtures. 
For a fermionic band-insulator, which can be assumed in the experimental realizations of Refs.~\cite{Ospelkaus2006, Gunter2006, Best2009, Heinze2011}, the boson-fermion interaction gives rise only to an irrelevant shift of the global chemical potential. Even for realistic assumptions for the confining potential, the interspecies interaction has little influence \cite{Kollath2004}. In contrast to the experimental results, the superfluid phase is even more stable within this framework. This poses the question of the applicability of the standard Hubbard model for quantum gas mixtures. It was pointed out that off-site interactions have a direct density-dependent influence on the total tunneling \cite{Hirsch1989, Strack1993, Hirsch1994, Amadon1996, Mazzarella2006, Mering2011, Luhmann2012, Bissbort2012} and that the inclusion of higher orbitals can have a strong impact on all Bose-Fermi Hubbard parameters, i.e., tunneling and on-site interactions \cite{Luhmann2008b, Lutchyn2009, Mering2011}.

\begin{figure*}
{\centering
\includegraphics[width=0.8\linewidth]{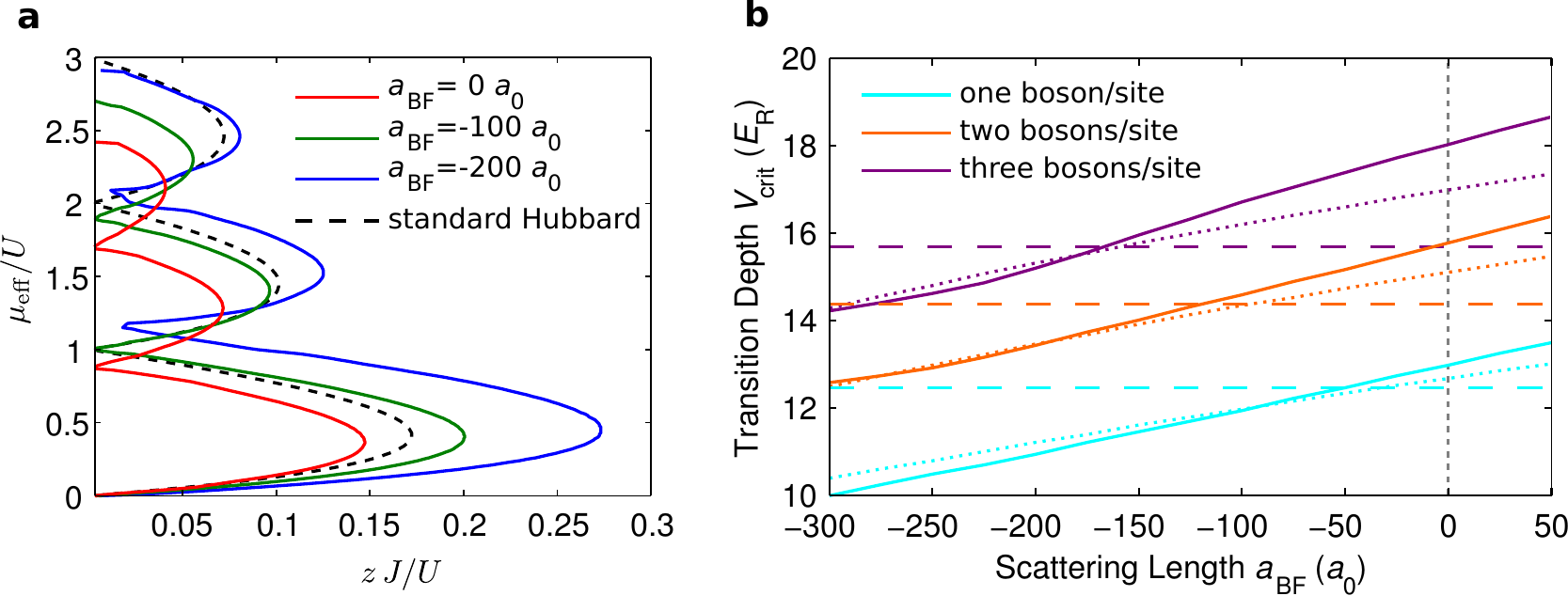}}
\caption{(Color online) \textbf{a} Phase diagram for the superfluid to Mott-insulator transition of bosons with a fermionic band-insulator at different interspecies attractions. The predictions of the standard Hubbard models are shown as a dahed black line. The attractive interaction effectively reduces the total tunneling resulting in extended Mott-lobes. \textbf{b} The critical lattice depth of the superfluid to Mott-insulator transition as a function of the interspecies scattering length $a_\BF$. The transition occurs at significantly shallower lattices than in the purely bosonic system $(a_\BF=0)$. The dashed lines correspond to the Bose-Fermi Hubbard model and the dotted lines to an extended model with only lowest-band processes (section \ref{sec:models}). 
\label{fig:PD}
}
\end{figure*}

In solid state systems, the impact of off-site interactions and higher orbitals has been addressed but is usually rather small \cite{Hirsch1989, Strack1993, Hirsch1994, Amadon1996}.
Unlike in solids, these effects can be significant in optical lattice systems due to the characteristic shape of the Wannier functions. However, they have been only recently discussed, e.g., in Refs.~\cite{Busch1998, Li2006, Mazzarella2006, Luhmann2008b, Lutchyn2009, Larson2009, Sakmann2009, Johnson2009, Sakmann2010, Hazzard2010, Will2010, Buchler2010, Dutta2011, Pilati2012}, and are mainly restricted to a single atomic species \cite{Busch1998, Li2006, Mazzarella2006, Larson2009, Sakmann2009, Johnson2009, Sakmann2010, Hazzard2010, Will2010, Buchler2010, Dutta2011, Pilati2012}. In general, either only multi-orbital effects \cite{Busch1998, Li2006, Luhmann2008b, Lutchyn2009, Larson2009, Sakmann2009, Johnson2009, Sakmann2010, Hazzard2010, Will2010, Buchler2010, Dutta2011, Pilati2012} \textit{or} off-site interactions \cite{Mazzarella2006} are covered. As elaborated in Refs.~\cite{Luhmann2012, Bissbort2012} for purely bosonic systems and Ref.~\cite{Mering2011}, the combination of both is essential for a correct 
description. The treatment of higher orbitals is usually performed only for few-site systems \cite{Busch1998, Sakmann2009, Sakmann2010} or by applying mean-field theory \cite{Li2006, Larson2009, Hazzard2010, Dutta2011}, which is insufficient in strongly correlated systems.

Recently, an  exact band-dressing method has been developed for single-component bosonic systems \cite{Luhmann2012, Bissbort2012}, which allows for an accurate treatment of higher-band processes. Here, we generalize this method to multi-component systems and in particular Bose-Fermi mixtures. The physical effects discussed in the following are in general present for all interacting quantum gas mixtures. The exact results, however, depend on the quantum statistics of the particles and the specific parameters such as the detuning from the light field and the atomic masses. For the example of a mixture of bosonic $\Rb$ and fermionic $\K$ we present accurate phase diagrams that significantly improve previous results. We find a large shift of the superfluid to Mott-insulator transition, which is considerably stronger than in reference \cite{Mering2011}. In contrast to our exact band-dressing method, the latter applies a band-elimination technique to treat orbital degrees of freedom.

As a central result, we present the phase diagram of the superfluid to Mott-insulator transition in a Bose-Fermi mixture in section \ref{sec:PD}. Furthermore, we discuss the crucial effect of off-site interactions and the corresponding shortcomings of the standard Bose-Fermi Hubbard model in section \ref{sec:off-site_interactions}. Subsequently, we present the procedure to incorporate higher-band processes in section \ref{sec:MO}. Afterwards, the corresponding extended Hubbard models and the implications for the bosonic phase transition are discussed in detail in section \ref{sec:models}.

\section{Phase diagrams}
\label{sec:PD}

We will now first discuss the resulting phase diagrams of the bosonic superfluid to Mott-insulator transition in the presence of a fermionic band-insulator, where the individual corrections to the standard Hubbard model are discussed in detail below. For concreteness, we choose a mixture of bosonic $\Rb$ and fermionic $\K$ in an optical lattice with a spacing of $a=377 \, \mathrm{nm}$ (experimental parameters of Ref.~\cite{Best2009}).
For the respective wavelength, the Wannier functions of both species are almost identical. The interaction between the bosonic atoms is fixed to a repulsive scattering length of $a_\BB=102\, a_0$ \cite{Will2010}, while the attractive interaction between the two species is tunable over a wide range using a Feshbach resonance \cite{Ferlaino2006, Best2009}. The fermionic nature of the spin-polarized potassium atoms simplifies the system, as we can assume a band-insulating phase and thus a fixed atom-number of one fermion per lattice site. This simplification is valid for experiments with high particle numbers and a strong confinement \cite{Kohl2005, Ospelkaus2006, Gunter2006, Schneider2008, Best2009, Heinze2011}. In principle, it is possible to directly apply the presented methods and extensions to systems where this assumption does not hold as well as to other quantum gas mixtures. In particular, both atomic species can be bosonic or fermionic and the generalization to multi-component systems with more than two species is straight forward.

For a fermionic band insulator, the fermionic degrees of freedom are frozen out and the physics can be described by an effective bosonic model that takes into account all effects induced by the interaction with the fermions. In the framework of this paper, we will discuss in detail the derivation of an effective Hamiltonian which reads
\begin{equation}
	\label{eq:Hext}
	\tilde H_\mathrm{ext}= -\sum_{\langle i,j \rangle}  \tilde{b}_i^\dagger \tilde{b}_j \tilde J^\tot_{\hat n_j,\hat n_i} 
	+ \sum_i \tilde E_{\hat{n}_i} - \mu \sum_i \hat{n}_i.
\end{equation}
We will see that despite its simplicity it already includes higher-band and bond-charge off-site processes. The latter gives rise to an occupation-dependent tunneling $J^\tot_{n_j,n_i}= J_\B + (n_i+ n_j-1) X_\BB + 2 X_\BF$ even within the lowest single-particle band. 
Here, $J_\B$ is the conventional tunneling; $X_\BB$ and $X_\BF$ are the bond-charge tunneling elements arising from Bose-Bose and Bose-Fermi interactions, respectively (section \ref{sec:off-site_interactions}). The interaction induced occupation of higher orbitals leads to a further occupation dependency of all parameters, i.e., $\tilde J_\B$, $\tilde X_\BB$, $\tilde X_\BF$ and $\tilde E_n$.

The tilde above the parameters and operators in \eqref{eq:Hext} indicates the multi-orbital dressing as discussed in section \ref{sec:MO}. The effective single-band Hamiltonian \eqref{eq:Hext} uses the ground state of the interacting system, called the dressed band, instead of the lowest single-particle band \cite{Luhmann2012,Bissbort2012}. The dressed operators $\tilde b_i$ and $\tilde b_i^\dagger$ annihilate and create bosonic particles on site $i$ in this dressed band and $\hat n_i=\tilde b_i^\dagger \tilde b_i$ counts the number of bosons on site $i$. It is important to note that after the transformation to the dressed band, the phase diagrams can be calculated using standard single-band methods. The renormalized on-site energy $\tilde E_n$ is composed of the single particle energies of bosons $\tilde \epsilon_{\mathrm{B},n}$ and fermions $\tilde \epsilon_{\mathrm{F},n}$ as well as the interaction energies for the repulsion between the bosons $\frac12 n(n-1) \tilde U_n$ and the attraction between the 
species $n \tilde U_{\mathrm{BF},n}$. The chemical potential $\mu$ fixes the total number of bosonic atoms.

After calculating the dressed parameters, we apply Gutzwiller mean-field theory to compute the critical lattice depth of the transition from the superfluid to the Mott-insulator. The phase diagrams of the extended model \eqref{eq:Hext} are shown in \fig{fig:PD}\textbf{a}. The effective chemical potential $\mu - \tilde E_1$ is given in units of the Hubbard on-site interaction $U$, where $\tilde E_1$ is the renormalized on-site energy of one boson and one fermion. For vanishing interaction between the bosons and fermions $a_{\BF}=0$ and a repulsive interaction $a_\BB=102\, a_0$ among the bosons, the Mott-lobes are contracted compared with the standard Hubbard model. This is a result of a decrease of the on-site energy and an increase of the total tunneling caused by off-site interactions \cite{Luhmann2012}. For increasing attraction between bosons and fermions the effect is reversed \cite{Mering2011} and the Mott-lobes are extended exhibiting a critical transition point at much lower lattice depths. This 
effect can be attributed to a strong reduction of the total tunneling amplitude induced by interspecies off-site interactions. 

In \fig{fig:PD}\textbf{b} the critical lattice depths for the superfluid to Mott insulator transition for one, two and three bosons per lattice site are shown as a function of the interspecies interaction strength. The solid lines depict the results obtained using the extended model \eqref{eq:Hext}, while the dashed lines correspond to the standard Bose-Fermi Hubbard model, which predicts no dependency on the interspecies interaction $a_\BF$. As discussed above, for $a_\BF=0$ the Mott-insulator transition is shifted to deeper lattices, where the shift is increased with the bosonic filling. This is mainly caused by the bosonic bond-charge interaction $\tilde X_\BB$ enhancing the total tunneling (section \ref{sec:off-site_interactions}). With increasing attractive interaction $a_\BF$ the transition is strongly shifted to shallower lattice depths.
Depending on the filling, the shift of the Mott-insulator transition caused by the fermionic atoms is $3$-$4\,\ER$ for $a_\BF=-300\, a_0$.
For a mixture of bosonic $\Rb$ and fermionic $\K$ the interspecies background scattering length is $a_\BF \approx -205\, a_0$ \cite{Ferlaino2006} and can be tuned by applying a Feshbach resonance \cite{Ferlaino2006, Best2009}.

The predicted shift is considerably larger than calculated with the adiabatic band elimination method in Ref.~\cite{Mering2011} which also incorporates the bond-charge interactions. The effective potential approach in Refs.~\cite{Luhmann2008b,Best2009} does not include the important contributions of fermionic on-site energy and bosonic bond-charge interaction. The extended Bose-Fermi-Hubbard model \eqref{eq:Hext} discussed here contains all relevant energies that can affect the superfluid to Mott-insulator transition at zero temperature. As a result, the Mott-insulator shift in Ref.~\cite{Best2009} can be partly explained by interaction-induced effects. This provides a consistent picture, where the experimental observations \cite{Ospelkaus2006,Gunter2006,Best2009} are a combined effect of the Hubbard extensions \textit{and} the adiabatic heating processes \cite{Cramer2008,Cramer2011} which depend on the initial temperature of the quantum gas.

\section{Off-site interactions}
\label{sec:off-site_interactions}


\begin{figure*}
{\centering
\includegraphics[width=0.985\linewidth]{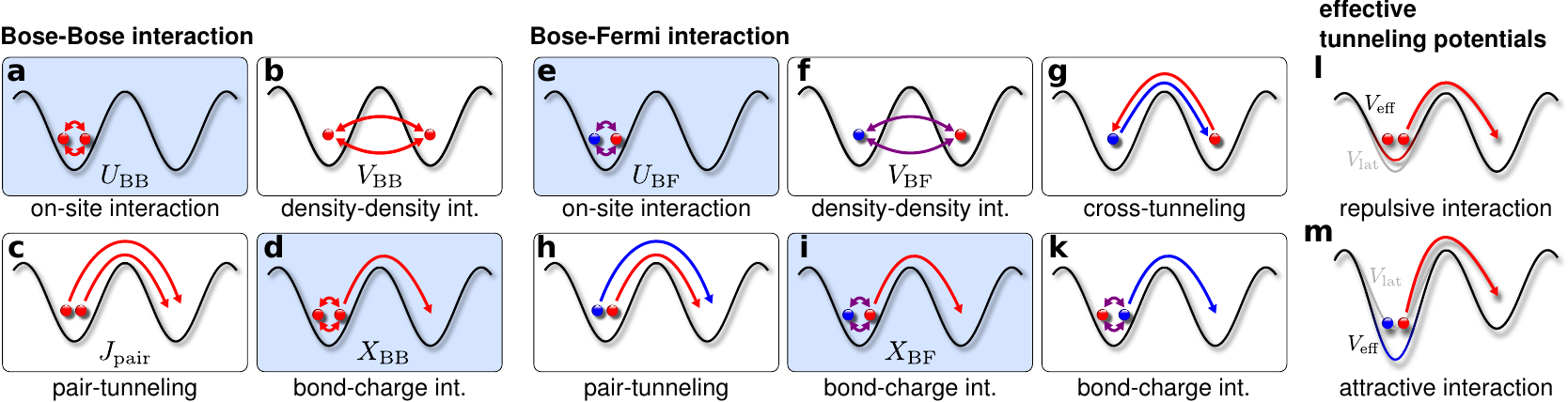}}
\caption{(Color online) Interaction-induced off-site processes. Processes induced by the Bose-Bose interaction are \textbf{a} bosonic on-site interaction, \textbf{b} density-density interaction, \textbf{c} pair-tunneling, and \textbf{d} bond-charge tunneling. Analogous interaction processes are induced by Bose-Fermi interaction, namely, \textbf{e} on-site interaction, \textbf{f} density-density-interaction, \textbf{g} cross-tunneling, \textbf{h} pair-tunneling, \textbf{i} bond-charge tunneling of a boson and \textbf{k} bond-charge tunneling of a fermion. The shading indicates the most important processes that are taken into account in the extended Hubbard model \eqref{eq:Hext}. The bond-charge tunneling processes (\textbf{d} and \textbf{i}) are illustrated in \textbf{l} and \textbf{m} via tunneling in effective potentials. 
\label{fig:offsite-processes}
}
\end{figure*}


We now turn back to the full description of the applied extended Hubbard model which features two corrections to the standard Hubbard model. First, off-site interactions lead to a significant contribution to the total tunneling amplitude by changing the effective tunneling potential. Second, the inclusion of multi-band processes causes a modification of all model parameters.

As mentioned above, mixtures of ultracold spin-polarized bosonic and fermionic atoms in optical lattices are usually described by the standard Bose-Fermi Hubbard model. The underlying tight-binding approximation restricts the model to the lowest single-particle orbital and interactions between particles on the same lattice site. The resulting Hubbard Hamiltonian reads
\begin{equation}\begin{split}
	 \hat H_\text{BFH}=& 	-  \sum_{\expect{i,j}} (J_\B \hat b_i^\dagger \hat b_j   
												\!+\! J_\F  \hat f_i^\dagger \hat f_j )
												+\frac{U_\BB}{2}  \sum_i \hat n_i (\hat n_i-1)  \\
										&		+\sum_i U_\BF \hat n_i \hat m_i  
												- \sum_i (\mu_\B \hat n_i + \mu_\F \hat m_i ),
\end{split}\label{eq:BFHM}\end{equation}
Here, $\hat b_i$ ($\hat f_i$) is the bosonic (fermionic) annihilation operator and $\hat n_i$ ($\hat m_i$) the respective particle number operator.
In general, the tunneling matrix elements for bosons ($J_\B$) and fermions ($J_\F$) can have different values. The on-site interaction is fully described by the parameters $U_\BB$ and $U_\BF$ for intra- and interspecies interaction, respectively. The total number of bosonic and fermionic atoms are fixed by the chemical potentials $\mu_\B$ and $\mu_\F$. Under common experimental conditions \cite{Ospelkaus2006,Gunter2006,Best2009,Heinze2011}, the fermions are in a band-insulator phase where Pauli-blocking prohibits tunneling. This freezes out the fermionic degrees of freedom and the resulting Hamiltonian captures the behavior of the bosons under the influence of exactly one fermion per lattice site. Consequently, we can set $\hat f_i^\dagger \hat f_j \rightarrow 0,\ \hat m_i \rightarrow 1$ and get
\begin{equation}\begin{split}
	 \hat H_\text{FBI}=& 	-  \sum_{\expect{i,j}} J_\B \hat b_i^\dagger \hat b_j   
												+\frac{U_\BB}{2}  \sum_i \hat n_i (\hat n_i-1)  \\
										&		+\sum_i (U_\BF-\mu_\B) \hat n_ i  .
\end{split}\label{eq:BHM}\end{equation}
The interaction energy $U_\BF$ between bosons and fermions can be absorbed into an effective chemical potential $\mu_\eff=\mu_\B-U_\BF$ and the resulting Hamiltonian does not differ from the standard Bose-Hubbard model. Thus, the behavior of the bosons is not influenced by the homogeneously distributed fermions, which is in contradiction to the experimental observations \cite{Ospelkaus2006,Gunter2006,Best2009}. 

In the derivation of the standard Hubbard model, it is argued that interaction processes between particles on neighboring lattice sites can be neglected due to their small amplitudes compared with on-site interactions. This argument is however only partly correct, since some of these processes involve the hopping of particles. In particular, the so-called bond-charge interactions (see \fig{fig:offsite-processes}) are of high relevance and will be discussed in the following. Compared to the conventional tunneling, these processes can be non-negligible and consequently alter the phase diagrams. In particular, the Wannier functions in optical lattices differ strongly from their counterpart in solid state materials leading to comparably large matrix elements for bond-charge processes. In addition, the possibility of larger fillings in bosonic systems can enlarge these interaction effects. Consequently, off-site interaction processes can be strongly enhanced for optical lattice systems. 

Consider the interacting part of the full two-particle Hamiltonian for the lowest band and two neighboring lattice sites $\mathrm{L}$ and $\mathrm{R}$  
 \begin{equation}
	  \hat H_\mathrm{int}= 
	\frac{1}{2} \sum_{ijkl}  U_{ijkl}^\BB\,
    \hat{b}^{\dagger}_i \hat{b}^{\dagger}_j \hat{b}_k \hat{b}_l + \sum_{ijkl}  U_{ijkl}^\BF\,
    \hat{b}^{\dagger}_i \hat{f}^{\dagger}_j \hat{f}_k \hat{b}_l,
	\label{eq:Hint}
\end{equation} 
with $i,j,k,l = \mathrm{L, R}$ and
\begin{equation}\begin{split}
   \label{eq:Uijkl_BB}
    U_{ijkl}^{\BB/\BF}  = {g_{\BB/\BF}} \int \ \
    &w^{\B *}_i\ofr\, w^{{\B/\F} *}_j(\mathbf{r'})\, V(\mathbf{r},\mathbf{r'})  \\ \times\,
  &w_k^{\B/\F}(\mathbf{r'})\, w_l^\B\ofr \ \ d^3r\, d^3r'.
\end{split}\end{equation}
The single-particle basis functions $w^{\B/\F}_i\ofr$ are the maximally localized Wannier functions, describing a boson/fermion sitting on site $i$. In general, the Wannier functions can be different for the individual atomic species depending on the atomic masses and detunings. The interaction strengths are given by $g_\BB = \frac{4\pi \hbar^2}{m_\B} \as$ and $g_\BF = \frac{2 \pi \hbar^2}{m_\mathrm{r}} a_\BF$ with the mass of the bosonic atoms $m_\B\ $ and the reduced mass $m_\mathrm{r}$ of boson and fermion. 
The interaction potential $V(\mathbf{r}, \mathbf{r'})$ we applied describes the scattering properties using a finite-ranged box potential (see \cite{Luhmann2012}). In a lowest-band treatment, this is usually replaced by contact interactions, i.e., a $\delta$-pseudopotential. In a multi-orbital framework the latter would lead to mathematical subtleties \cite{Busch1998}.


\begin{figure}[b]
{\centering
\includegraphics[width=0.8\linewidth]{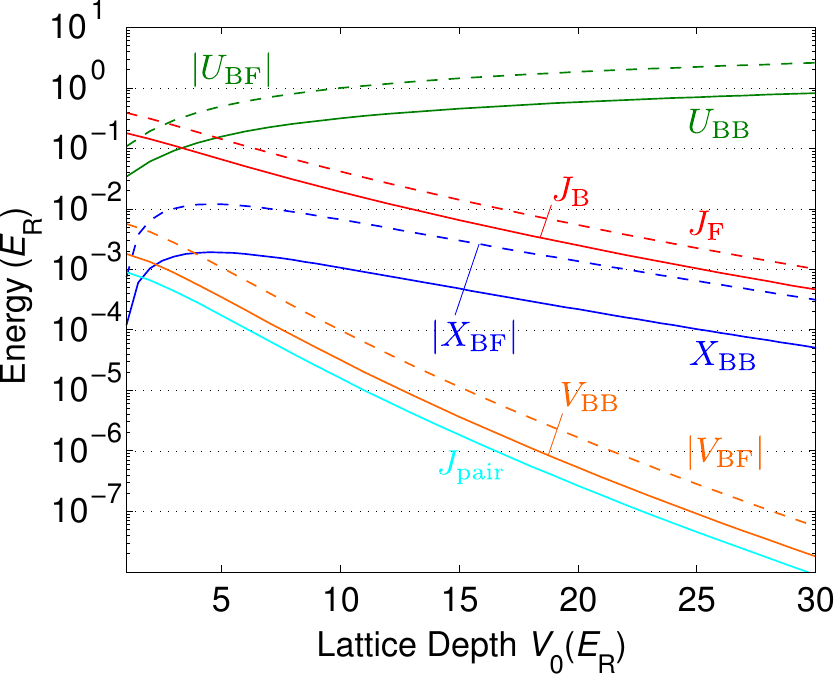}}
\caption{(Color online) Matrix elements for the different hopping and interaction processes. The solid lines are the purely bosonic processes, while the dashed lines correspond to Bose-Fermi interaction and fermionic tunneling.  Shown are the intra- and interspecies on-site interaction $U_\BB$ and $U_\BF$ (green), the conventional tunneling amplitudes $J_\B$ and $J_\F$ (red), the bosonic bond-charge tunneling $X_\BB$ and $X_\BF$ (blue) as well as density-density interactions $V_\BB$ and $V_\BF$ (orange) and correlated pair-tunneling $P_\BB$ (cyan). The amplitudes are calculated for scattering lengths $a_\BB = 102\, a_0$ and $a_\BF = -200 \, a_0$.
\label{fig:ProcessIntegrals}
}
\end{figure}


The distinct processes arising from the full two-body Hamiltonian \eqref{eq:Hint} for Bose-Bose and the Bose-Fermi interaction are depicted in \fig{fig:offsite-processes}. In addition to the on-site interaction (\fig{fig:offsite-processes}\textbf{a}), the Bose-Bose interaction leads to the density-density interaction process $V_\BB\, \hat n_i \hat n_j$ (\fig{fig:offsite-processes}\textbf{b}), the correlated tunneling of a particle pair  $P_\BB\, \hat b_i^{\dagger2} \hat b_j^{2}$ (\fig{fig:offsite-processes}\textbf{c}), and the bond-charge assisted tunneling $-X_\BB\, \hat b_i^{\dagger} (\hat n_i+ \hat n_j) \hat b_j$ (\fig{fig:offsite-processes}\textbf{d}). The respective matrix elements are $V_\BB=U^\BB_{ijji}$, $P_\BB=U^\BB_{iijj}/2$, and $X_\BB=-U^\BB_{iiij}=-U^\BB_{ijjj}$. 
The amplitudes of these processes are plotted in \fig{fig:ProcessIntegrals} as solid lines. While the density-density interaction $V_\BB$ and the pair tunneling amplitude $P_\BB$ are several orders of magnitudes smaller than the the standard Hubbard processes, the bond-charge interaction $X_\BB$ is only one order of magnitude smaller than the conventional tunneling $J_\B$. As the bond-charge interaction $-X_\BB\, \hat b_i^{\dagger} (\hat n_i+ \hat n_j) \hat b_j$ scales with the particle number on both involved sites, it can easily reach non-negligible values and must be accounted for (see also \cite{Luhmann2012}). In addition, all interaction processes scale linearly with the interaction strength which can be tuned experimentally. Here, it is essential that the bond-charge interaction contributes to the tunneling of the particles rather than the on-site interaction. These estimations hold for a wide range of lattice depths $V_0 \gtrsim 5\, \ER$ and scattering lengths $|a_\BB| \gtrsim 100\, a_0$.

For the Bose-Fermi interaction more distinct processes exist as the interacting particles are distinguishable.
First, the cross-tunneling (\fig{fig:offsite-processes}\textbf{g}), which is similar to the density-density interaction (\fig{fig:offsite-processes}\textbf{f}) except that the particles interchange, and second, the bond-charge interaction, where either a bosonic (\fig{fig:offsite-processes}\textbf{i}) or a fermionic (\fig{fig:offsite-processes}\textbf{k}) particle tunnels. However, assuming a fermionic band insulator all processes involving the hopping of a fermion are forbidden. The respective amplitudes of the processes induced by Bose-Fermi interaction are plotted as dashed lines  in \fig{fig:ProcessIntegrals}.

In conclusion, only on-site interactions and bond-charge tunneling of bosons have to be taken into account (shaded in \fig{fig:offsite-processes}), since other processes are prohibited by the fermionic band-insulator or contribute only with extremely small amplitudes. Thus, the necessary extensions of the Hubbard Hamiltonian, i.e. the bond-charge processes in \fig{fig:offsite-processes}\textbf{d} and \textbf{i}, read
\begin{equation}\begin{split}
\hat X = \hat X_\BB + \hat X_\BF = \hat b_i^\dagger (&U_{iiij}^\BB\hat b_i^\dagger \hat b_i + U_{ijjj}^\BB\hat b_j^\dagger \hat b_j \\
+ &U_{iiij}^\BF\hat f_i^\dagger \hat f_i + U_{ijjj}^\BF\hat f_j^\dagger \hat f_j) \hat b_j.
\label{eq:Xsb}
\end{split}\end{equation}

An intuitive physical understanding of the bond-charge tunneling induced by Bose-Bose as well as by Bose-Fermi interaction can be obtained by the analogy to an effective tunneling potential \cite{Luhmann2012}. Assuming contact interactions in the single-band description and using the integral expressions \eqref{eq:Uijkl_BB}, we can rewrite the expression \eqref{eq:Xsb} as
\begin{equation}
\label{eq:bond-charge_operator}
 \hat{X} = \hat b_i^\dagger \hat b_j \dr w_i^{\B*}\ofr \left(g_\BB \hat \rho_{ij}^\BB + g_\BF \hat \rho_{ij}^\BF \right) w_j^\B\ofr,
\end{equation}
where we introduced the reduced densities
\begin{align}
 \hat \rho_{ij}^\BB &= \hat n_i |w_i^\B\ofr|^2 + (\hat n_j-1) |w_j^\B\ofr|^2 \\
 \hat \rho_{ij}^\BF &= \hat m_i |w_i^\F\ofr|^2 + \hat m_j |w_j^\F\ofr|^2.
\end{align}
The $-1$ in the bosonic density corresponds to the exclusion of self-interactions and is directly obtained from the commutation relations.
Inside the integral, we can replace these operators by the density functions $\hat \rho_{ij}^\BB \rightarrow \rho_\B\ofr - |w_j^\B\ofr|^2$ and $\hat \rho_{ij}^\BF \rightarrow \rho_\F\ofr$. In contrast to the operators $\hat\rho_{ij}$, the functions $\rho\ofr$ contain the density of particles on distant sites ($k\neq i, j$). However, this does not contribute to the integral \eqref{eq:bond-charge_operator} due to the small overlap with the Wannier functions $w_i\ofr$ and $w_j\ofr$. The bond-charge tunneling operator \eqref{eq:bond-charge_operator} can now easily be unified with the conventional tunneling to find the expression
\begin{equation}
\hat J + \hat X = \bra{w_i^\B\ofr} \left(\frac{\mathbf{p}^2}{2m} + V_\eff\ofr \right) \ket{w_j^\B\ofr} \hat b_i^\dagger \hat b_j
\end{equation}
which corresponds to the conventional tunneling in an effective potential $V_\eff\ofr = V\ofr + g_\BB(\rho_\B\ofr - |w_j^\B\ofr|^2) + g_\BF \rho_\F\ofr$. 

In \fig{fig:offsite-processes}\textbf{k} and \textbf{l} the tunneling in effective potentials is sketched.
Repulsive interactions, as between the bosons, effectively reduce the lattice depth, while attractive interactions have the opposite effect. Depending on the relative scattering lengths, the total tunneling can be strongly enhanced or reduced. This is consistent with the results for the fermionic tunneling obtained in \cite{Heinze2011}. The modification of the total tunneling in the lowest band already leads to a considerable deformation of the well-known phase diagram of the superfluid to Mott-insulator transition for bosons as discussed later in detail.

\section{Multi-orbital renormalization}
\label{sec:MO}

Whereas in the previous section we introduced off-site interactions as an important extension to the standard Hubbard model, we will now discuss another important feature of the Hamiltonian \eqref{eq:Hext}, namely the effective inclusion of higher bands. In the standard Hubbard model approach, only the lowest single-particle band is assumed to be occupied. However, in the strongly correlated system particles are promoted to higher orbitals due to the interaction-induced coupling between the orbitals. By changing their wave functions the particles minimize their on-site interaction energy (see section \ref{sec:MO}.B).

As a result of the population of higher orbitals, the effective wave function overlap of particles on neighboring lattice sites changes. This leads to modified amplitudes of the tunneling and the off-site interactions. The orbital occupation itself, however, is solely determined by on-site interactions due to their dominating contribution to the total energy. Consequently, the single-site problem can be handled separately (see section \ref{sec:MO}). 
Note that off-site processes are negligible for the determination of the orbital occupations but not for the calculation of the phase diagrams as elaborated in sections \ref{sec:PD} and \ref{sec:off-site_interactions}.

\subsection{Multi-orbital dressing}
The approach presented in the following can be divided into four steps.
(i) First, the single-site many-particle problem is solved using the method of exact diagonalization, which is described in section \ref{sec:MO}.B. The respective ground state $\Psi(n)$ is a fully-correlated many-particle state of $n$ bosons containing the full information on the population of higher orbitals. 
(ii) From the ground state $\Psi(n)$, we construct a multi-orbitally \textit{dressed} band with creation and annihilation operators that fulfill the usual relations  $\tilde b_i \ket{\Psi(n)}_i = \sqrt{n} \ket{\Psi(n-1)}_i$ and $\tilde b_i^\dagger \ket{\Psi(n)}_i = \sqrt{n+1} \ket{\Psi(n+1)}_i$. For example, a tunneling processes in the dressed band is represented by $\tilde b_i^\dagger \tilde b_j$.
(iii) In the next step, the amplitudes of, e.g., tunneling and bond-charge interactions in the dressed band are calculated. These amplitudes effectively contain processes in all orbitals and are explicitly occupation-number dependent.
(iv) Finally, we obtain an effective single-band Hamiltonian, where the dressed band replaces the lowest Bloch band. This dressed-band model now allows to apply standard single-band methods to calculate the phase diagram (ranging from mean-field to quantum Monte-Carlo methods). In the following, we give a detailed description of the steps (i) to (iii) while the last step (iv) is elaborated in sections \ref{sec:PD} and \ref{sec:models}.

(i) From the calculations in section \ref{sec:MO}.B we obtain the single-site ground state, which is a superposition of many-particle states $\Psi(n)=\sum_{N, M} c_{N,M} \ket{N}\ket{M}$.
Here, $\ket{N}\ket{M}=\ket{n_0,n_1,...}\ket{m_0,m_1,...}$ is the product state with $n_\alpha$ bosons and $m_\alpha$ fermions in the Wannier orbital $w_{\B/\F}^\a \ofr$, where $\alpha$ indicates the orbital and $n = \sum_\alpha n_\alpha$. In the following we will assume a fermionic band-insulator with $m=\sum_\alpha m_\alpha = 1$. The state $\Psi(n)$ consists of the lowest single-particle band dressed with small contributions of higher bands and we will refer to it in the following as the dressed band.
At zero temperature the particles will exclusively occupy the many-particle state $\Psi(n)$ instead of the lowest single-particle band.

(ii) Now we turn to the representation of operators in the dressed band using the ground state $\Psi(n)$.
Within a lowest-band treatment a general two-site operator can be decomposed in operators of the form
\begin{equation}
\label{eq:SO-operator}
 \hat O_\mathrm{lowest-band} = A\ \hat O_\L\, \hat O_\R,
\end{equation}
with an amplitude $A$ and operators $O_i$ consisting of creation/annihilation operators $\hat b_i^\dagger\,/\,\hat b_i$ on the left (L) or right (R) site. As an example, for the tunneling operator from the right to the left site it is $A=-J$, $\hat O_\L = \hat b^\dagger_\L$, and $\hat O_\R=\hat b_\R$. 

The corresponding operator of the dressed band $\tilde O_i$ can be directly obtained from $\hat O_i$ by replacing the operators $\hat b_i^\dagger\,/\,\hat b_i$ with their dressed counter-parts $\tilde b_i^\dagger\,/\,\tilde b_i$. The effective two-site operator $\tilde O$ in the dressed band takes the form
\begin{equation}
\tilde O = \tilde A_{\nL,\nR} \tilde O_\L \tilde O_\R,
\end{equation}
with the occupation-number dependent amplitude $\tilde A_{\nL,\nR}$.

(iii) In order to calculate the dressed-band amplitude $\tilde A_{\nL,\nR}$ that effectively includes all orbital processes, we start with the multi-orbital two-site operator. It can be decomposed in the same way as for the lowest-band
\begin{equation}
\label{eq:MO-operator}
 \hat O = \sum_{\{\alpha\},\{ \beta\}} A^{\{\alpha\},\{ \beta\}} \hat O_\L^{\{\alpha\}} \hat O_\R^{\{\beta\}},
\end{equation}
where the summation is over all possible sets of orbitals $\{\alpha\}=\{\alpha_1,\alpha_2,... \}$ and $\{\beta\}=\{\beta_1,\beta_2,... \}$. $A^{\{\alpha\},\{ \beta\}}$ is the amplitude for the corresponding process and $\hat O_i^{\{\alpha\}}$ consists of creation and annihilation operators $\hbd{$i$}{\alpha_k}$ and $\hb{$i$}{\alpha_k}$ for particles on site $i$ in the orbital $\alpha_k$. In the simplest case -- the conventional tunneling -- we have only sets with a single orbital $\{\alpha\}=\alpha$ and $\{\beta\}=\beta$,  the operators on the left and the right site $\hat O_\L^{\alpha} = \hbd{L}{\alpha}$, $\hat O_\R^{\beta} = \hb{R}{\beta}$, and tunneling amplitudes $A^{\alpha,\beta}$ between the orbitals $\alpha$ and $\beta$ as defined below. 

The effective amplitude $\tilde A_{\nL,\nR}$ is obtained from the matrix element $\bra{\Psi_\F} \hat O \ket{\Psi_\I}$, where $\Psi_\I(\nL,\nR)$ denotes the initial and $\Psi_\F=\Psi(\nL',\nR')$ the final state of the process. It thereby includes the summation over all multi-orbital processes. Since the states are product states of the individual lattice sites $\ket{\Psi(\nL)}\ket{\Psi(\nR)}$, also the effective amplitude $\tilde A$ decomposes into individual site contributions
\begin{equation}\begin{split}
 \tilde A_{\nL,\nR} \!= \!\frac{1}{N} \!\sum_{\{\alpha\},\{ \beta\}} \!\! A^{\{\alpha\},\{ \beta\}} &\bra{\Psi(\nL')} \hat O_\L^{\{\alpha\}} \ket{\Psi(\nL)} \\ 
\times & \bra{\Psi(\nR')} \hat O_\R^{\{\beta\}} \ket{\Psi(\nR)} ,
\end{split}\end{equation}
where $N=\bra{\Psi_\F} \tilde O_\L \tilde O_\R \ket{\Psi_\I}$ is needed for the correct normalization. Note that the effective amplitude is intrinsically occupation-dependent. 


\begin{figure}
{\centering
\includegraphics[width=0.8\linewidth]{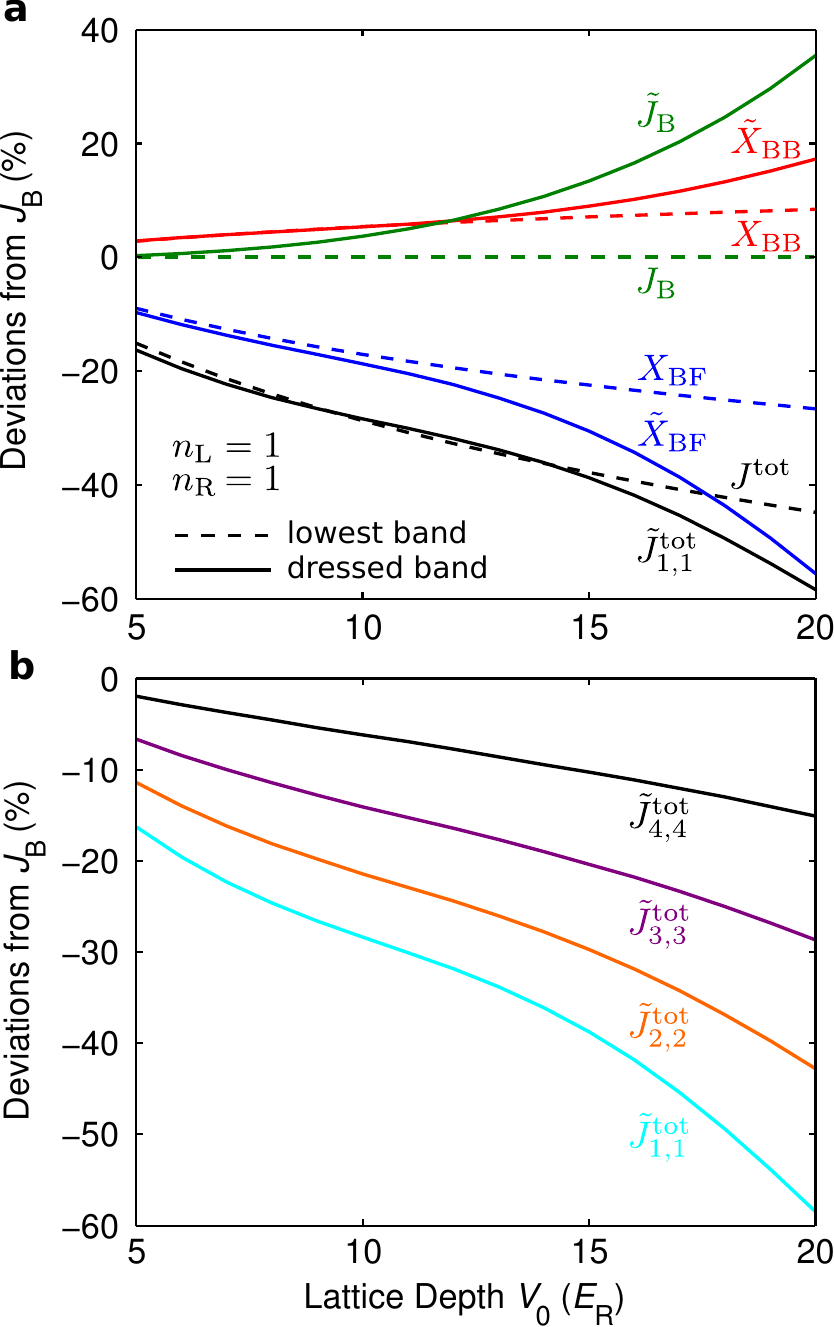}}
\caption{(Color online) \textbf{a} Contributions to the total tunneling in the lowest band (dashed) and the dressed band (solid) for one bosons and one fermion on each site. The total tunneling (black) is composed of conventional tunneling $J$ (green), bosonic bond-charge tunneling $X_\BB$ (red), and fermionic bond-charge tunneling $X_\BF$ (blue). The absolute value of all matrix elements are enhanced by orbital degrees of freedom but due to their opposite signs the multi-orbital effects partly compensate each other. \textbf{b} Deviations of the total tunneling $\tilde{J}^\mathrm{tot}_{n,n} = \tilde J_\B + (n_\L+n_\R-1)\tilde X_\BB + 2 \tilde X_\BF$ from the standard Hubbard tunneling $J_\B$ for different bosonic occupations $n_\L=n_\R$.
\label{fig:tunneling_contributions}
}
\end{figure}


As an example, for the conventional single-particle tunneling of bosons it follows $A^{\alpha,\beta}=-J_{\alpha} \delta_{\alpha,\beta}$ and $N=\sqrt{\nL}\sqrt{\nR+1}$. Here, $J_{\alpha}=-\bra{w^{(\alpha)}} \frac{\mathbf{p}^2}{2m} + V\ofr \ket{w^{(\alpha)}}$ is the tunneling amplitude in band $\alpha$. 
As another example, the multi-orbital bosonic bond-charge operator
\begin{equation}\begin{split}
\hat X_\BB = \ \sum_{\{\alpha\},\{\beta\}} &X_\BB^{\{\alpha\},\{ \beta\}}\, \hbd{L}{\alpha_1}
\hbd{L}{\alpha_2}\hb{L}{\alpha_3}\hb{R}{\beta_1}+\\
\sum_{\{\alpha'\},\{\beta'\}} &X_\BB^{\{\alpha'\},\{ \beta'\}}\, \hbd{L}{\alpha_1'}
\hbd{R}{\beta_1'}\hb{R}{\beta_2'}\hb{R}{\beta_3'}
\end{split}\end{equation} 
decomposes in left and right part $\hat O_\L^{\{\alpha\}}$ and $\hat O_\R^{\{\beta\}}$, that consist of either one or three creation/annihilation operators. In contrast to the conventional tunneling, orbital-changing processes are allowed for the multi-orbital bond-charge operator. 

The renormalized amplitudes are depicted in \fig{fig:tunneling_contributions} for all relevant contributions to the total tunneling in the case of one boson and one fermion per site (solid lines).
The amplitudes for conventional tunneling $\tilde J_{n_\L, n_\R}$ (green line) as well as bond-charge induced tunneling $\tilde X_{\BB,n_\L, n_\R}$ (red) and $\tilde X_{\BF,n_\L, n_\R}$ (blue) can differ strongly from the lowest band values, which are indicated as dashed lines. The multi-orbital renormalization can enhance the conventional tunneling by up to $30\%$ and the bond-charge induced processes can even be twice as strong as in the lowest single-particle band.
In the present situation of repulsive bosons and attractive Bose-Fermi interactions, the effect is strongest for low boson numbers (see \fig{fig:tunneling_contributions}\textbf{b}). This is mainly caused by a compensation of Bose-Bose and Bose-Fermi interaction induced bond-charge tunneling. Processes with imbalanced boson numbers can be calculated analogously  and are similarly affected. Due to symmetry reasons the relation $J_{\nL,\nR}=J_{\nR+1, \nL-1}$ holds for all tunneling processes from left to right.

\subsection{On-site problem and on-site interaction}
This section is dedicated to the explicit solution of the many-body problem on a single lattice site for a given number of particles. We apply the method of exact diagonalization to compute the ground state $\Psi(n)$ of $n$ bosons and $m=1$ fermion on a single lattice site. In particular, this leads directly to an occupation-number-dependent on-site energy $E_n$. The Hamiltonian of the single-site problem reads
\begin{equation}\begin{split}
	\label{eq:Hsite}
	\hat{H}_\text{site}=&\sum_\alpha \epsilon_\B^\a \hat{n}^\a + \sum_\alpha \epsilon_\F^\a \hat{m}^\a \\
+ \frac12 &\sum_{\abcd} U_\BB^{(\abcd)}\, \hbd{}{\alpha}\hbd{}{\beta}\hb{}{\gamma}\hb{}{\delta} \\
+ &\sum_{\abcd} U_\BF^{(\abcd)}\, \hbd{}{\alpha}\hfd{}{\beta}\hf{}{\gamma}\hb{}{\delta},
\end{split}\end{equation}
with the bosonic particle number operator $\hat n^\a=\hbd{}{\alpha}\hb{}{\alpha}$ and the single-particle energies $\epsilon_\B^\a$. The operators $\hat m^\a$, $\hfd{}{\alpha}$, $\hf{}{\alpha}$ and the energy $\epsilon_\F^\a$ are the fermionic analogues. The multi-band interaction amplitudes are defined as
\begin{equation}\begin{split}
    U_{\BB/\BF}^{(\abcd)}  =  {g_{\BB/\BF}} \int \  \ 
    &w^{\a*}_\B\ofr\, w^{(\beta)*}_{\B/\F}(\mathbf{r'})\,  V(\mathbf{r},\mathbf{r'}) \\ \times\,
 &w^{(\gamma)}_{\B/\F}(\mathbf{r'})\, w^{(\delta)}_\B(\mathbf{r}) \ \ d^3r \, d^3r',
\end{split}\end{equation}
where the interaction potential $V(\mathbf{r},\mathbf{r'})$ is a finite-range box potential with a width of $7.5\,\text{nm}$ (see Ref.~\cite{Luhmann2012} for details). These processes reflect the interaction-induced transition of particles from the orbitals $\gamma$ and $\delta$ to the orbitals $\alpha$ and $\beta$. This couples the different orbitals, resulting in a correlated multi-orbital ground state. We expand this Hamiltonian in the basis of many-particle Fock-states $\ket{N}\ket{M}$ and apply exact diagonalization, the so-called configuration interaction method. We restrict the calculation to the lowest 9 bands per spatial direction and use a high energy cutoff \footnote{The resulting many-body product basis has a total length of $12000 n^2$, where $n$ is the number of bosons.}. The on-site energy is directly obtained as the lowest eigenvalue of the matrix and its contributions can be computed as expectation values of the individual operators in \eqref{eq:Hsite} using the corresponding eigenvector. 

\begin{figure}
{\centering
\includegraphics[width=0.8\linewidth]{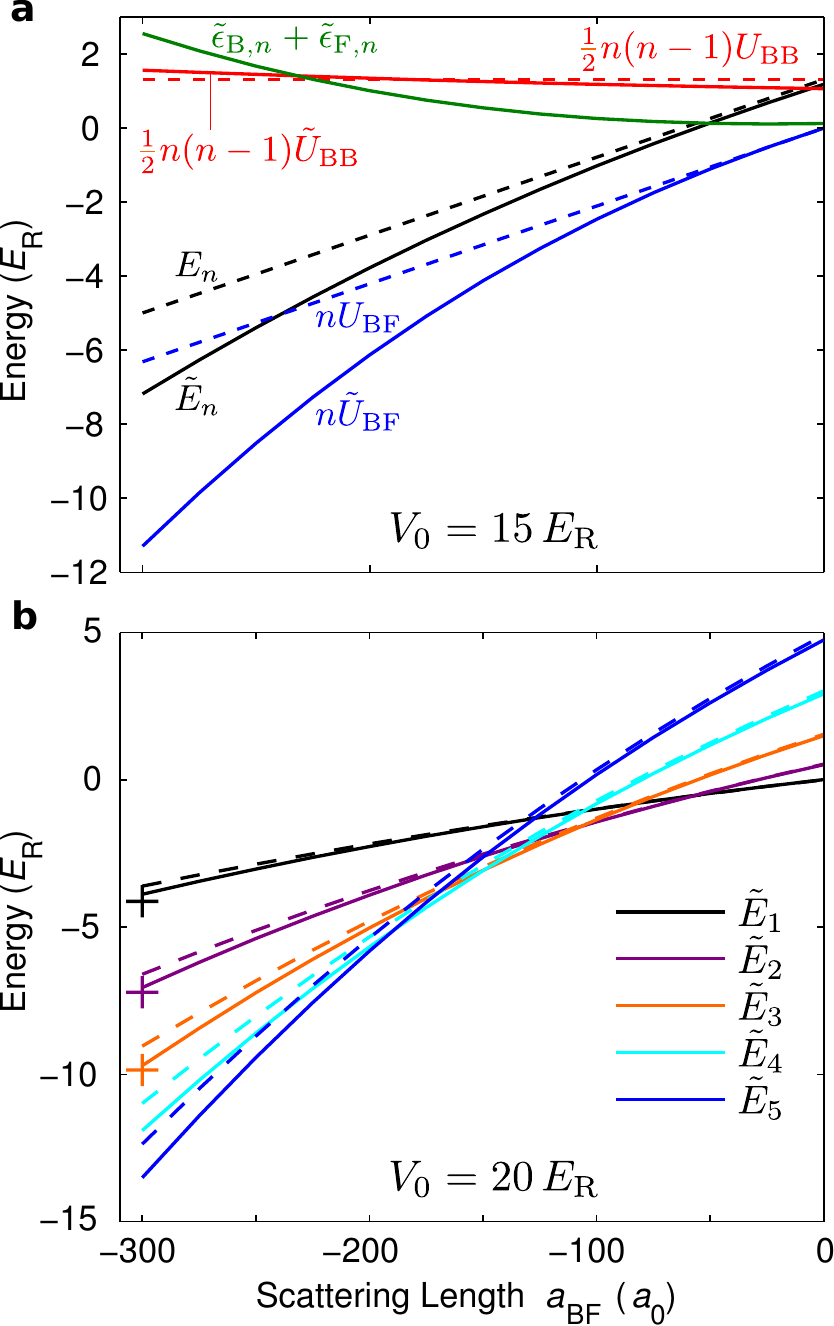}}
\caption{(Color online) \textbf{a} Total on-site energy $E_n$ and its contributions for $n=3$ bosons and $m=1$ fermion as functions of the interspecies scattering length. The contributions are single-particle energies $\epsilon_{\B,\F}$ (green), Bose-Bose interaction $U_\BB$ (red), and Bose-Fermi interaction $U_\BF$ (blue). The total on-site energy is visibly lower than the lowest band prediction (dashed lines).  \textbf{b} The total on-site energies $E_n$ for various numbers of bosons $n$ and one fermion. With increasing interspecies interaction the total energy decreases non-linearly. The dashed lines correspond to a three times larger interaction range. The markers are obtained by applying scaling theory as described in the text.  
\label{fig:onsite_contributions}
}
\end{figure}

The energy contributions are plotted in \fig{fig:onsite_contributions}\textbf{a} for $V_0=15\,\ER$ and $n=3$ bosons (solid lines). The values significantly deviate from the lowest-band approximation (dashed lines). The single-particle energies $n \tilde \epsilon_\B + \tilde \epsilon_\F$ (green line) are measured relative to the lowest-band values and thus are always positive. The occupation of higher orbitals causes a contraction of the wave functions which leads to an increase of the absolute value of the Bose-Fermi interaction. This results in a large reduction of the total on-site energy (black line) for large scattering lengths $\asBF$. Of course, also the repulsive interaction among the bosonic particles (red line) contributes but is less drastically influenced. 


The total on-site energies for various boson numbers are shown in \fig{fig:onsite_contributions}\textbf{b}. The dashed lines correspond to a calculation with a three times larger interaction range. In general, the contributions of higher bands are reduced with an increasing interaction range. The figure shows that the energy is only weakly affected by a the change of the (finite) interaction range.  Additionally, we applied scaling theory to estimate the value of convergence for the on-site energy at the numerically most demanding parameters $V_0=20\,\ER$ and $a_\BF=-300\,a_0$ for an interaction range of $7.5\,\text{nm}$. We see that the error (differences in energies) converges exponentially with both the length of the many-particle basis and  the number of orbitals.  Scaling our results according to the exponential behavior, we are able to determine the converged energy value, where we first perform the scaling in respect to the basis length for a given number of orbitals. The results indicated as crosses 
in \fig{fig:onsite_contributions}\textbf{b} show only small deviations and justify the constraints for the basis length and the number of orbitals applied for the solid lines.

The total on-site energy becomes intrinsically occupation-dependent (beyond the dependency in the standard Bose-Fermi Hubbard model  \eqref{eq:BFHM}) and can be written in terms of effective n-particle collisions \cite{Will2010, Johnson2009,Bissbort2012} 
\begin{equation}
\tilde E_n=n \tilde E_1 + \frac12 n(n-1) \bar E_2 + \frac{1}{6} n(n-1)(n-2) \bar E_3\text{...}
\end{equation} 
The first term $\tilde E_1$ describes the interaction energy by two-particle collisions between bosons and fermions. The second term is the interaction energy caused by processes that involve two bosons $\bar{E}_2=\tilde E_2-2 \tilde{E}_1$ and the third term $\bar{E}_3=\tilde E_3 - 3\bar{E}_2 - 3 \tilde{E}_1$ involves three bosons. Although, the restriction to the first three terms is enough to describe well the energies for up to $n=5$ bosons, we use here the exact values for $\tilde E_n$.

\section{Superfluid to Mott-insulator transition}
\label{sec:models}

\begin{figure}
{\centering
\includegraphics[width=0.8\linewidth]{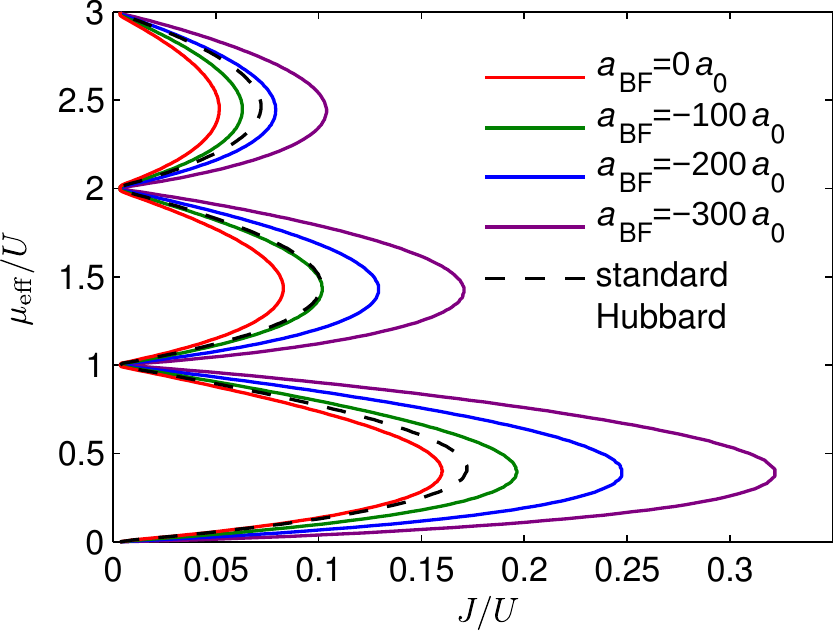}}
\caption{(Color online) Phase diagrams of the lowest-band model \eqref{eq:extended_BFHM}  with bond-charge interactions for different scattering lengths. For comparison the results of the standard Hubbard model are shown as a dashed black line.
\label{fig:PD_SB}
}
\end{figure}

After discussing off-site interactions and multi-orbital renormalizations, we now turn back to the full many-body quantum gas problem. Obviously, from the above results, it is necessary to take both bond-charge interactions and higher bands into account. First, we now define an extended model of the lowest band that includes off-site interactions and discuss its implications. Afterwards, we replace the lowest single-particle band and parameters with the respective dressed analogues and thereby include higher bands in a very efficient way. 

The extended Hubbard model of the lowest band reads
\begin{equation}\begin{split}
 \hat H_\text{ext}=& - \sum_{\expect{i,j}} \big[J_\B\!+\! X_\BB (\hat n_i + \hat n_j - 1) \!+\! 2\, X_\BF \big]\hat b_i^\dagger \hat b_j\\
&+\frac{U_\BB}{2}  \sum_i \hat n_i (\hat n_i-1) - \mu_\eff \sum_i\hat n_i.
\end{split}\label{eq:extended_BFHM}\end{equation}
While the repulsive interaction between the bosons increases the total tunneling, the attractive fermions reduce the bosonic mobility. As one central result and in strong contrast to the predictions of the standard Hubbard model (see section \ref{sec:off-site_interactions}) the superfluid to Mott-insulator transition is shifted. The phase diagrams are shown in \fig{fig:PD_SB} for different attractive Bose-Fermi interaction strengths. For strong Bose-Fermi attraction and low bosonic filling, the transition occurs at much shallower lattices, due to the effectively deepened tunneling potential. The effect is reversed when the repulsion between the bosons becomes stronger than the attraction to the fermions, which is the case for weaker Bose-Fermi interaction and higher bosonic filling. In \fig{fig:PD}\textbf{b} the critical lattice depth for the transition is plotted as a function of the interspecies interaction strength $a_\BF$. The dotted lines correspond to the extended Hamiltonian \eqref{eq:extended_BFHM} 
restricted to the lowest 
band.

When including higher bands we must replace the lowest-band operators with those of the dressed band $\tilde b_i$ and $\tilde b_i^\dagger$. Also, the parameters $J$, $X$ and $U$ must be renormalized as discussed in section \ref{sec:MO}. All tunneling contributions, i.e., conventional tunneling, both bond-charge interactions as well as their multi-orbital renormalizations can be combined to one total tunneling parameter 
\begin{equation}
\tilde J^\tot_{n_i, n_j}= \tilde J_{n_i,n_j}+\tilde X_{\BB,n_i,n_j}(n_i + n_j -1)+2 \tilde X_{\BF,n_i,n_j},
\end{equation} 
which is explicitly occupation-number dependent. The renormalized on-site energy 
\begin{equation}
\tilde E_n = n \tilde \epsilon_{\mathrm{B},n} + \tilde \epsilon_{\mathrm{F},n} + \frac12 n(n-1) \tilde U_n + n \tilde U_{\mathrm{BF},n}
\end{equation} 
is composed of the renormalized single particle energies of bosons $\tilde \epsilon_{\mathrm{B},n}$ and fermions $\tilde \epsilon_{\mathrm{F},n}$ as well as the interaction energies for the repulsion between the bosons $\frac12 n(n-1) \tilde U_n$ and the attraction between the species $n \tilde U_{\mathrm{BF},n}$. This allows to define the extended Hamiltonian of the dressed band in equation \eqref{eq:Hext}, namely,
\begin{equation}
	\tilde H_\mathrm{ext}= -\sum_{\langle i,j \rangle}  \tilde{b}_i^\dagger \tilde{b}_j \tilde J^\tot_{\hat n_j,\hat n_i} 
	+ \sum_i \tilde E_{\hat{n}_i} - \mu \sum_i \hat{n}_i.
\end{equation}
The dressed-band Hamiltonian now takes into account all higher-band processes and all relevant nearest-neighbor interactions. The multi-orbital corrections of the Bose-Fermi interaction have a strong impact on the chemical potential at which the transition to a certain Mott-lobe occurs. This distorts and shifts the phase diagram along the axis of the chemical potential. Therefore, we plot \fig{fig:PD}\textbf{a} in terms of an effective chemical potential $\mu_\eff = \mu - \tilde E_{1}$. Concerning the transition point of the bosonic superfluid to Mott-insulator transition, the reduction of the total on-site energy by multi-orbital processes counters the effect of reduced total tunneling. Nonetheless, the total effect on the transition can be a shift of several recoil energies $\ER$ depending on interaction strengths and filling factors (\fig{fig:PD}\textbf{b}).  Note that the lowest-band model $\hat H_\mathrm{ext}$ underestimates the impact on 
the superfluid to Mott-
insulator transition by only up to $1 \ER$, which is surprising keeping the strong changes of the individual amplitudes in mind. However, this (coincidental) compensation of the contributing amplitudes depends on the choice of system parameters. Furthermore, it has been demonstrated that the on-site interaction \cite{Will2010,Will2011} and the effective tunneling matrix element \cite{Heinze2011} are experimentally well accessible and can be measured independently.  

\section{Density-density interactions and pair-tunneling}
\label{sec:other_processes}
In the above model, several off-site processes have been neglected due to their small amplitude in the lowest band. In this context the question arises, whether the multi-orbital dressing can enhance them to non-negligible values. The correlated pair tunneling and the density-density interactions both have very small amplitudes because in the integrand the small tail of the Wannier function enters quadratically on both lattice sites. By contrast,  in the bond-charge integral the tail is multiplied three times with the center of a Wannier function. However, when taking into account strongly delocalized wave functions of higher bands, this argument no longer applies and all processes need to be reconsidered, since the overlap integrals become comparable for all types of off-site interactions.  
These contributions are strongly suppressed in the case of density-density interactions, because the initial and final state both depend on higher-band contributions with small coefficients $c_N \ll 1$ to produce a large integral, whereas in the case of pair tunneling one of them can be the ground state $(c_N \approx 1)$. Although, the multi-orbitally renormalized pair tunneling is usually smaller than the conventional tunneling it can, in general, reach the same order of magnitude. Due to this structure it has very bad convergence properties in respect to the total number of contributing bands. As a fourth order contribution in perturbation theory it can, however, be neglected even for rather large amplitudes. Density-density interactions are not as strongly influenced and remain small. This justifies, the restriction to the extensions incorporated in Hamiltonian \eqref{eq:Hext}, which includes all relevant off-site interactions.

\section{Conclusions}
\label{sec:conclusions}
We have discussed the important role of interaction effects in atomic quantum gas mixtures in optical lattices. Off-site interactions as well as higher band processes turn out to have a strong impact on these systems, which we are able to calculate using an extended occupation-dependent Hubbard model. In particular, we have focused on Bose-Fermi mixtures in this paper, where the standard Bose-Fermi-Hubbard model fails to cover all relevant processes. This manifests itself in a strong shift of the superfluid to Mott-insulator transition in the bosonic subsystem, which is not predicted by the standard Hubbard model. The critical lattice depth is shifted towards shallower lattices with increasing Bose-Fermi attraction. Similar corrections are present for all experiments with optical lattices and can be expected to be relevant, e.g., for Bose-Bose mixtures, low-dimensional systems or other lattice geometries. Omitting the condition of having a fermionic band insulator, which we have applied here, the presented 
extensions of the Bose-Fermi Hubbard model can lead to very rich physics such as the formation of polarons.

We have shown that for optical lattice systems the bond-charge tunneling  \cite{Hirsch1989,Strack1993,Hirsch1994,Amadon1996,Mazzarella2006,Mering2011,Luhmann2012,Bissbort2012} is the most important contribution of the nearest-neighbor off-site interaction as it can drastically influence the tunneling. Repulsive interactions enhance the total tunneling, whereas the attractive interactions reduce it. In an intuitive picture, this can be described as lowered and increased effective potentials, respectively. Furthermore, higher band processes not only reduce the total on-site energy but also have an impact on the conventional tunneling and the bond-charge interactions. We have treated the problem by dressing \cite{Luhmann2012,Bissbort2012} the lowest single-particle band with interaction-induced occupations of higher-orbital states. This leads to a renormalization of interactions and tunneling parameters that become intrinsically occupation-dependent. These parameters have been used in order to define an 
extended Bose-Fermi-Hubbard model capable of describing effects of higher orbitals and off-site interactions in Bose-Fermi mixtures.

The results show in general that interactions in multi-component systems can have a crucial impact beyond the standard Hubbard treatment. In the presented case, the standard Hubbard model is incapable of describing the interspecies interaction between bosonic and fermionic atoms correctly. While here mainly the effects on the bosonic atoms have been discussed, the mutual interaction affects the fermionic atoms similarly, which has recently been observed in experiment \cite{Heinze2011}.

\section{Acknowledgments}
We thank U. Bissbort and W. Hofstetter for stimulating discussions and acknowledge financial support by the German science foundation DFG under grant FOR801.


\end{document}